\documentclass[12pt]{article}

\usepackage{scicite}

\usepackage{times}
\usepackage{graphicx}
\usepackage{amsmath}
\usepackage{amssymb}
\usepackage{verbatim}
\usepackage{color}
\usepackage{lscape} 
\usepackage{multirow}
\usepackage{booktabs} 

\newenvironment{sciabstract}{%
\begin{quote} \bf}
{\end{quote}}

\newcounter{lastnote}

\title{Is Europe Evolving Toward an Integrated Research Area?}

\author
{Alessandro Chessa,$^{1}$ Andrea Morescalchi,$^{1}$ Fabio Pammolli,$^{1}$$^\ast$ \\
Orion Penner,$^{1}$ Alexander M. Petersen,$^{1}$ Massimo Riccaboni,$^{1}$$^\ast$\\
\\
\normalsize{$^{1}$ IMT Lucca Institute for Advanced Studies,  Lucca  55100, Italy}\\
\\
\normalsize{ Published in {\it Science} {\bf 339}, 650-651 (2013).} \\
\normalsize{$^\ast$To whom correspondence should be addressed; }\\
\normalsize{E-mail:  f.pammolli@imtlucca.it. or massimo.riccaboni@imtlucca.it}
}

\begin{document}


\baselineskip24pt

\maketitle

\begin{sciabstract}
An integrated European Research Area (ERA) is a critical component for a more competitive and open European R\&D system.
However, the impact of EU-specific integration policies aimed at overcoming innovation barriers associated with national borders is not well understood.
Here we analyze $2.4 \times 10^{6}$ patent applications filed with the European Patent Office (EPO) over the 25-year period 1986--2010 along with a sample of $2.6\times10^{5}$ records from the ISI Web of Science to quantitatively measure the role of borders in international R\&D collaboration and mobility. From these data we construct five different networks for each year analyzed: (i) the patent co-inventor network, (ii) the publication co-author network, (iii) the co-applicant patent network, (iv) the patent citation network, and (v) the patent mobility network.
We use methods from network science and econometrics to perform a comparative analysis across time and between EU and non-EU countries to determine the ``treatment effect'' resulting from EU integration policies. Using non-EU countries as a control set, we  provide quantitative evidence that, despite decades of efforts to build a European Research Area, there has been little integration above global trends in patenting and publication. This analysis provides  concrete evidence that Europe remains a collection of national innovation systems.
\end{sciabstract}

Efforts towards European research and development (R\&D) integration have a long history, intensifying with the Fifth Framework Program (FP) in 1998 \cite{Delanghe2009,ScherngellBarber2011,IntegrationERA} and the launch of the European Research Area (ERA) initiative at the Lisbon European Council in 2000. A key component of the European Union (EU) strategy for innovation and growth \cite{EuropePolicyForum,ERA1} is the ERA aims at an integrated innovation system through directed funding, increased mobility, and streamlined innovation policies that can overcome national borders.

To assess the rate of progress towards this ERA vision, we analyze the evolution of geographical collaboration networks constructed from patent and scientific publication data. While these data may not capture every facet of ERA, they are widely accepted measures of R\&D output and the European Commission considers them crucial for the evaluation of the Horizon 2020 FP \cite{ImpactAssessmentReport}. All in all, we find no evidence since 2003 that EU innovation policies aimed at promoting an integrated research and innovation system have corresponded to intensified cross-border R\&D activity in Europe vis-\`a-vis other OECD countries. 

We exploit the June 2012 release of the OECD REGPAT database \cite{MarautDernisWebbSpieziaGuellec}, and analyze all $\sim 2.4\times 10^{6}$ patent applications filed with the European Patent Office (EPO) over the period 1986-2010. For comparison with scientific publications we take a random sample of $\sim 2.6\times 10^{5}$ records from the ISI Web of Science over the period 1991-2009. We geo-coded each data set at the NUTS3 region level (see Supplementary Materials (SM)). 
Using the data we construct 5 networks, which provide different perspectives into EU R\&D integration. In our networks, nodes correspond to NUTS3 regions and links represent collaboration/mobility measures. Specifically, (i) the patent co-inventor network and (ii) the publication co-author network measure the intensity of inter-regional collaboration at the individual level; (iii) the co-applicant patent network measures the collaboration between institutions (``applicants'') located in different regions; (iv) the patent citation network indirectly measures scientific integration by following the flow of citations from patents in one region to patents in another; (v) the patent mobility network measures the mobility of inventors from one region to another by tracking their location in subsequent patents.

\begin{figure*}
\centering{\includegraphics[width=0.99\textwidth]{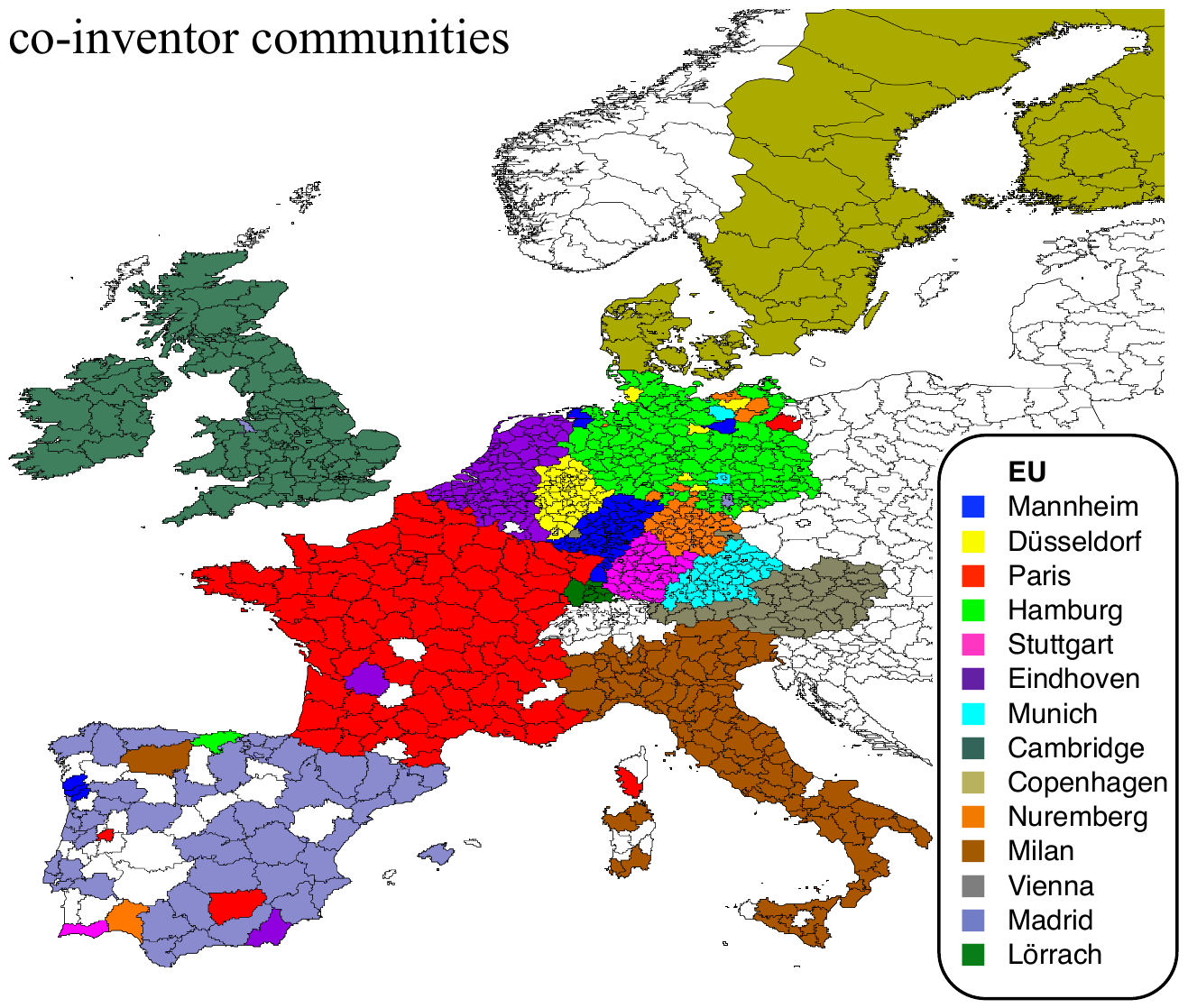}}
\caption{The community structure of the 2009 EU-15 co-inventor network. Communities are shown with different colors and are labeled by their most central region. Communities have been generated by iteratively aggregating nodes (NUTS3 regions) into clusters of increasing size (see SM). Blank regions have no ties in 2009.}
\label{Fig1}
\end{figure*}

We use a standard network-clustering algorithm to identify communities, i.e., sub- sets of nodes more strongly linked to one another than to nodes outside, to compare geopolitical borders and R\&D networks. Regional integration is shown in Fig. \ref{Fig1} in the purple community, centered on Eindhoven, which is composed of strongly collaborating regions in the Benelux, and in the international Nordic community with its center in Copenhagen. Despite these exceptions, patterns of co-inventorship in Europe continue to be largely shaped by national borders. This observation stands in contrast to the community structure of the highly dispersed ``coast-to-coast'' US co-inventor network (see SM for comparison) \cite{UnivIndustry}.
Figure \ref{Fig1} shows Europe as a collection of regional and national innovation communities. However, that does not necessarily mean that integration efforts have been unsuccessful. The more relevant question, then, is at what rate is Europe evolving toward an integrated research system relative to the rate of cross-border R\&D collaboration observed in non-EU OECD countries.

Consistent with recent studies \cite{IntegrationERA,Breschi2004,HoekmanScherngellFrenkenTijssen2012,NSB} we observe a significant increase in the total number of cross-border research collaborations, both within and outside Europe (see SM). To assess the role of EU-specific factors, we compare the relative change in cross-border collaboration between European countries (e.g., distinguishing German-French, from German-German, and French-French collaborations) to the relative change in cross-border collaboration between non-European OECD countries (e.g., distinguishing USA-Japan from USA-USA, and Japan-Japan collaborations). Collaborations between EU and non-EU regions are not included in our analysis.

\begin{figure*}
\centering{\includegraphics[width=0.99\textwidth]{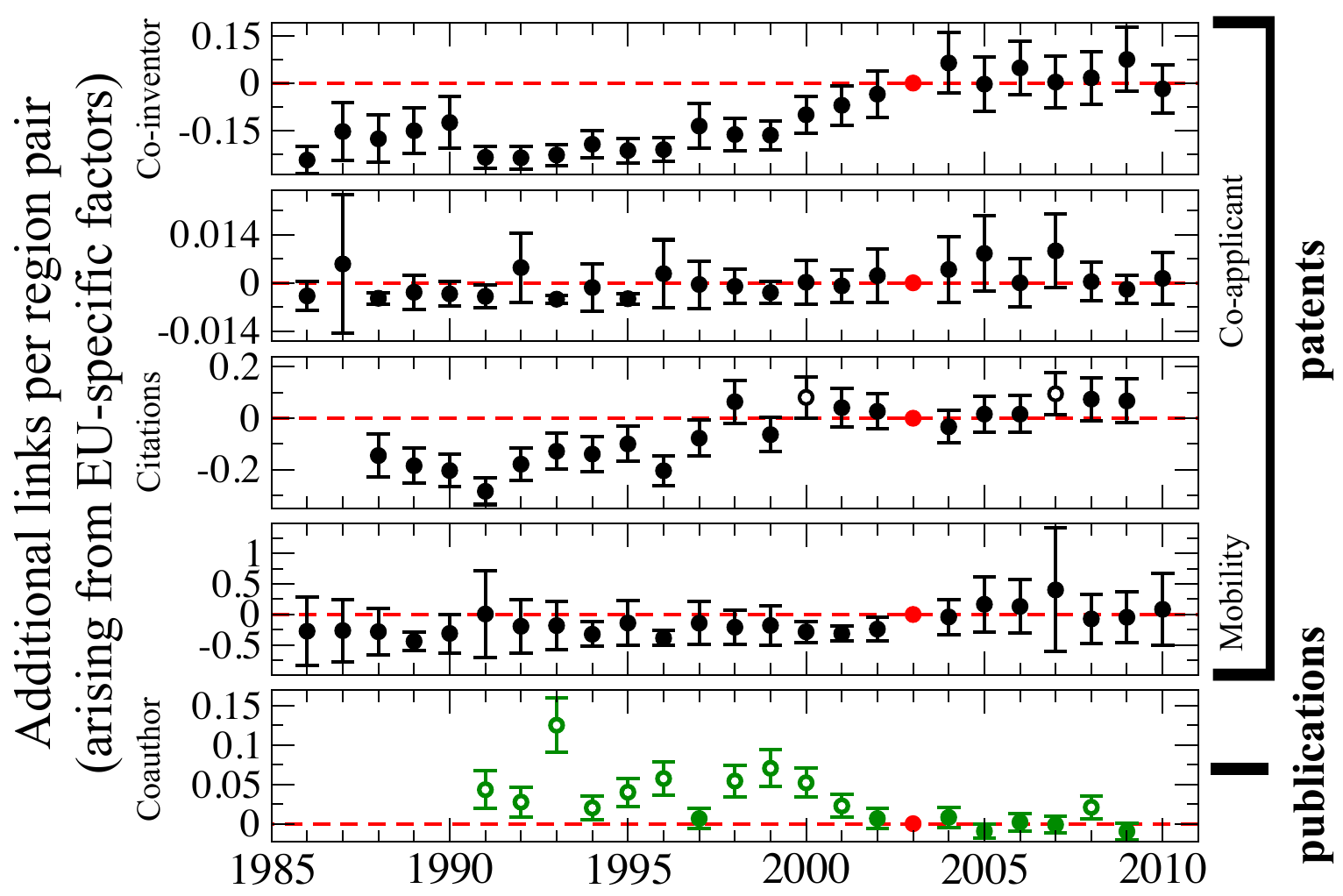}}
\caption{ The evolution of European integration in five R\&D networks. We use econometric methods to measure the effect of EU specific factors on the amount of cross-border links relative to within-border links and to the rest of non-EU OECD countries. Results are shown for 4 different patent networks (black circles) and a scientific publication network (green circles). Open circles indicate statistically significant (.05 level) positive deviations from the baseline year (2003). The y-axis reports the additional number of links for an average pair or regions relative to 2003 due to R\&D integration in Europe.}
\label{Fig2}
\end{figure*}

For each network, our econometric model simultaneously performs three quantitative differences and controls for the size of regions, geographic distance and time effects (see SM). First, the difference between cross-border and intra-border average number of links is computed, both for EU and non-EU OECD nations. Second, the difference between these two estimates isolates the impact of EU-specific factors on R\&D integration. The final one, to a baseline year, yields the quantitative output of the model, i.e. the expected number of additional links between regions resulting from EU specific factors. This quantity is shown in Figure \ref{Fig2}. Comparing data points from two different years, a higher y-axis value indicates a greater impact of EU specific factors upon integration among EU nations. Thus by construction, choice of the baseline year does not alter our results. It also follows that a positive (negative) slope indicates Europe is integrating faster (slower) than non-EU countries.

Since the late 1990s, we observe some signs of integration in European patent statistics. In the case of the patent co-inventor network, there has been an increase in cross-border collaboration in Europe vis-\`a-vis other OECD countries. This effect was relatively pronounced from 1998 to 2002, but stalled in 2003. Since then, the additional number of links for an average pair of regions due to Europe specific factors has never been significantly larger than zero. The patent co-applicant network exhibits no significant increase since 1996. The citation network shows a temporary bump in integration in the late 90's, then fluctuates around that level. Finally, the inventors' mobility network shows almost no progress in the last decade, confirming a slow pace of integration for the European high-skill labor market.

The scientific publications co-authorship network shows a negative trend since 1999, indicating that cross-border links among non-EU OECD countries grew faster than European cross-border links. These results are striking and deserve further investigation given the amount of resources the EU has committed to promote cross-border scientific collaboration through programs like FPs, European Cooperation in Science and Technology, Networks of Excellence, Marie Curie Actions, etc.

In sum, our analysis of R\&D patent and publication networks shows Europe remains a collection of loosely coupled national innovation systems \cite{nelson1993national}. Furthermore, since 2003, cross-border collaborations in Europe have developed no faster than in the rest of the OECD countries. 
Several ongoing initiatives seek to address a number of general shortcomings that have affected previous integration efforts \cite{ERA1}. The European Institute of Innovation and Technology's (EIT) Knowledge and Innovation Communities are long-term (7-15 years) collaboration networks spanning all aspects of the R\&D ecosystem \cite{eit}. To foster synergetic interaction between national funding bodies, Science Europe, an association of national research organizations, was founded in 2011 \cite{scienceeurope}. 

The European Research Council \cite{erceuropa} has taken major steps to promote cross-border mobility by making grants competitive and portable. Likewise, a memorandum of understanding signed by the European Commission and the League of European Research Universities \cite{leru} pushes for pension unification and transparency in hiring and tenure decisions.

Despite these initiatives to increase competition within the system, monitoring and evaluation must drastically change if Europe is to accomplish its ambitious goals in Science and Technology. Evidence based evaluation focused on output and impact is crucial, as recognized in the plans for the Horizon 2020 FP \cite{ImpactAssessmentReport}. Our methodology promotes this vision by combining interdisciplinary expertise with data relevant to evaluation. 

\section*{Acknowledgements}
Authors acknowledge funding from the National Research Program of Italy (PNR) project ÒCRISIS LabÓ and thank M. Buchanan and F. Cerina for comments. M.R. acknowledges funding from the Italian Ministry of Education, University, and Research (MIUR) [Project of National Interest (PRIN) project 2009Z3E2BF] and the Research Foundation of Flanders (FWO) (G073013N). O.P. acknowledges funding from the Social Sciences and Humanities Research Council of Canada.

\newpage
\clearpage

\begin{center}
\section*{Supplementary Materials\\}
\end{center}

\bigskip
\begin{center}
{\bf \large Is Europe Evolving Toward an Integrated Research Area?} \\
Alessandro Chessa,$^{1}$ Andrea Morescalchi,$^{1}$ Fabio Pammolli,$^{1}$ \\
Orion Penner,$^{1}$ Alexander M. Petersen,$^{1}$ Massimo Riccaboni,$^{1}$\\
$^{1}$IMT Institute for Advanced Studies Lucca, 55100 Lucca, Italy \\
(Feb. 11, 2013)
\end{center}
\bigskip
\renewcommand{\theequation}{S\arabic{equation}
}
\renewcommand{\thefigure}{S\arabic{figure}}
\renewcommand{\thetable}{S\arabic{table}}

\setcounter{equation}
{0}  
\setcounter{figure}{0}
\setcounter{table}{0}
\setcounter{section}{0}

\section{Materials and Methods} 

We perform a geo-spatial network analysis of scientific collaboration. In our framework, the nodes are NUTS3 regions.\footnote{The Nomenclature of Units for Territorial Statistics (NUTS) is a geo-code standard for referencing the subdivisions of countries for statistical purposes. The nomenclature has been introduced by the European Union, for its member states. The OECD provides an extended versions of NUTS3 for its non-EU member and partner states.}
In our analysis of R\&D integration, we distinguish between two types of collaboration links: (a) links between NUTS3 within the same country, and (b) cross-border links between NUTS3 regions in country $m$ and NUTS3 regions in country $n$, as demonstrated by the green links in Fig. \ref{FigS1}(A). Fig. \ref{FigS1}(B) outlines our methodological approach where we analyze and compare the time evolution of collaboration networks in EU countries
vis-\`a-vis non-EU countries. We use an econometric 
model to measure the difference  between the  network structure in
year  $t^{*}+\Delta t$ and  the ``baseline year'',  which we choose to be $t^{*}=2003$.

Supplementary materials are organized as follows: the first section describes our data sources and database construction; the second section illustrates the network clustering methods we employed; and the third section contains a detailed description of our statistical methodology and results.  All relevant Data and Code can be downloaded at: \\{\tt http://cse.lab.imtlucca.it/SOM/SOM.zip}

\subsection{Data}

Patent collaboration data are drawn from the OECD REGPAT database \cite{MarautDernisWebbSpieziaGuellec,REGPAT}
which compiles all patent applications filed with the European Patent Office (EPO) since the 1960s. Within this database the geographical location of inventor and applicant
are designated by one of the 5,552 NUTS3 regions in 50 countries.\footnote{In our analysis we considered 40 countries. European countries consists of the EU-15: Austria, Belgium, Germany, Denmark, Spain, Finland, France, United Kingdom, Greece, Ireland, Italy, Luxembourg, Netherlands, Portugal, Sweden. The control set is comprised of 25 other nations outside of the EU-15: Australia, Bulgaria, Brazil, Canada, Switzerland, Chile, China, Hong Kong, Croatia, Israel, India, Iceland, Japan, South Korea, Liechtenstein, Macedonia, Mexico, Norway, New Zealand, Romania, Russian Federation, Turkey, Taiwan, United States, South Africa. EU-15 Gross Domestic Expenditures on R\&D (GERD) spending is 2.6\% of their combined GDP. In the case of the non-EU control set, that number is 2.1\%. Likewise, the distribution of GERD within each set is similar, with a mix of high and low spending countries. On average for EU-15 countries, 15\% of the R\&D budget comes from the EU and the remaining 85\% from national budget. Conversely, while statistical figures are not available the shared R\&D budget in non-EU OECD countries is considerably smaller.}
We use all patent applications across all classes in the REGPAT database over the period 1986--2010, $~2.4\times10^{6}$ applications overall.\footnote{Data for 2010 might be incomplete as some EPO filings are published with lags and may not appear in the data yet.}
We construct 4 geographical networks: (i) co-inventors, (ii) co-applicants, (iii) citations and (iv) inventor mobility.
In (i) and (ii) the strength of a link between two regions is equal to the number of patents jointly invented by or jointly assigned to the two regions. In (iii) it is the number of patent citations between inventors' regions. Specifically, for each pair $(i, j)$ of NUTS3 regions we count the number of times that (a patent invented by an inventor residing in) region $i$ cites (a patent invented by an inventor residing in) region $j$. Conversely the number of citations that $i$ receives from $j$ is the strength of the link $(j, i)$. In (iv) link weight is equal to the number of inventors moving from region $i$ to region $j$. As for citations, the mobility network is directed, i.e. we distinguish between mobility from $i$ to $j$ and mobility in the opposite direction, $j$ to $i$. Links are created tracking regional migration for inventors with at least two patents. We compare the affiliation of inventors' consecutive patents and assign a new link whenever a new patent is filed in a region different from the one reported in the inventor's previous patent. \footnote{OECD REGPAT database provides a unique identifier for inventors' name. For more detailed information on patenting activity the reader can refer to a survey of inventors for around 9,000 European patented inventions (35).}

\begin{figure*}
\centering{\includegraphics[width=0.8\textwidth]{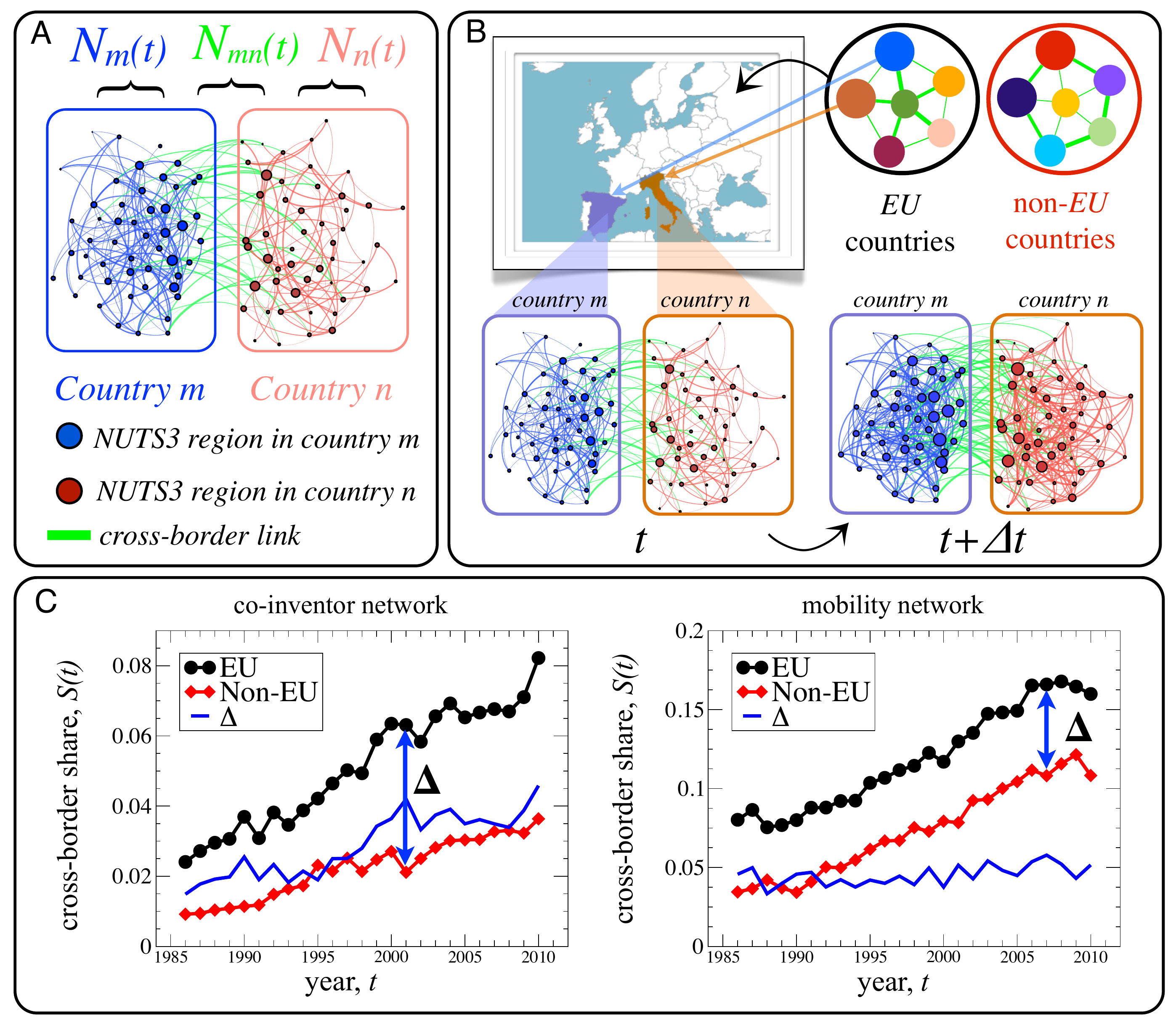}}
\caption{Network analysis of co-patent activity. (A) Schematic illustration of the network methodology. For each year $t$ we calculate weighted links from $N_{n}(t)$, the number of patents between NUTS3 regions within a country $n$, and $N_{mn}(t)$, the number of patents between NUTS3 regions in different countries, as indicated by the green links. (B) The evolution of the collaboration networks over time serves as the basis for analyzing the integration rate of the EU innovation system, and these within-EU changes over time are compared to non-EU changes over time. (C) For the set of EU countries, we show the annual cross-border share $S(t) = N_{\times}/N_{T}$, calculated as the ratio of the number $N_{\times}(t) = \sum_{m,n} N_{mn}$ of cross-border collaboration links divided by the total number $N_{T}(t) = N_{\times} + \sum_{n} N_{n}$ of both intra- and cross-border collaboration links. We calculate the same quantity for the set of non-EU countries. The increase of S over time in the co-inventor and mobility networks reflects a well-documented increasing trend in global patent activity. However, the share difference $\Delta = S(EU)-S(nonEU)$, a coarse indicator of relative integration that does not control for EU specific factors, is relatively flat for both measures, except for a small ``jump'' around 1998-2000 in the co-inventor network. The relatively constant trend in $\Delta(t)$ is preliminary empirical evidence that brings into question the effectiveness of EU policies aimed at accelerating integration. Our econometric ``treatment effect'' approach further investigates the effectiveness of EU integration policies by controlling for multiple underlying variables, see Eq. S5. Link count values  are listed in Table \ref{linkcounts}.}
\label{FigS1}
\end{figure*}

Fig. \ref{FigS1}(C) illustrates the global trend of increased cross-border
collaboration in the co-inventor network, and increased cross-border
flow in the mobility network. We count the total number of
intra-border collaborations links  $N_{i}=\sum_{n} N_{n}(t)$ and the
total number of cross-border collaboration links
$N_{\times}=\sum_{m,n}N_{m,n}$ and define the collaboration share to
be the ratio $S\equiv N_{\times}/(N_{i}+N_{\times}$) for a given time
period (values shown in Fig. \ref{FigS1}(C) and values for the other networks are listed in Table \ref{linkcounts}). The overall increasing
trend reflects both the increasing pace of patenting and the decreasing
role of distance in worldwide research efforts.
We note that for the case of Europe the 15\% final share matches the
ratio of the EU research budget to the combined national research
budgets of EU nations.

Scientific publications data are drawn from ISI-Web of Science. The Web of Science database is a bibliographical collection maintained by Thomson Reuters, considered to be one of the most comprehensive and reliable sources of information on research activity across all fields of science. We analyze a random sample of 256,015 research articles in the period 1991-2009 by authors affiliated at institutions located in the OECD countries. We build the regional co-authorship network by geo-coding each address attached to the paper. Since addresses refer to institutional affiliations and it is not possible to link individuals to organizations  \cite{PondsVanOortFrenken2007}, we define co-authorship as the co-occurrence of two or more addresses on a publication. Therefore, if an author lists multiple affiliations in different regions we consider co-authorship links between those regions in our analysis.

\subsection{Community detection}
 There are now many community detection methods for clustering networks
 \cite{Fortunato}, one the most popular being modularity optimization,
 introduced by Newman and Girvan
 \cite{NGcommunity}. Some limitations have been noted for this method, the most
 important being the existence of a resolution limit \cite{FortunatoB}
 that prevents it from detecting small
 modules. Nevertheless it is reliable for  standard cluster analysis provided a suitable optimization procedure is employed. In the
 present analysis we adopt a weighted version of
 the modularity function and optimized it using the Louvain algorithm
 \cite{Blondel}. This algorithm arrives at the final community
 structure by starting from isolated nodes
 (NUTS-3 regions in our case) and iteratively aggregating them into
 communities of increasing size. This particular optimization procedure
 can mitigate the effect of the resolution limit.

\begin{figure*}
\centering{\includegraphics[width=0.99\textwidth]{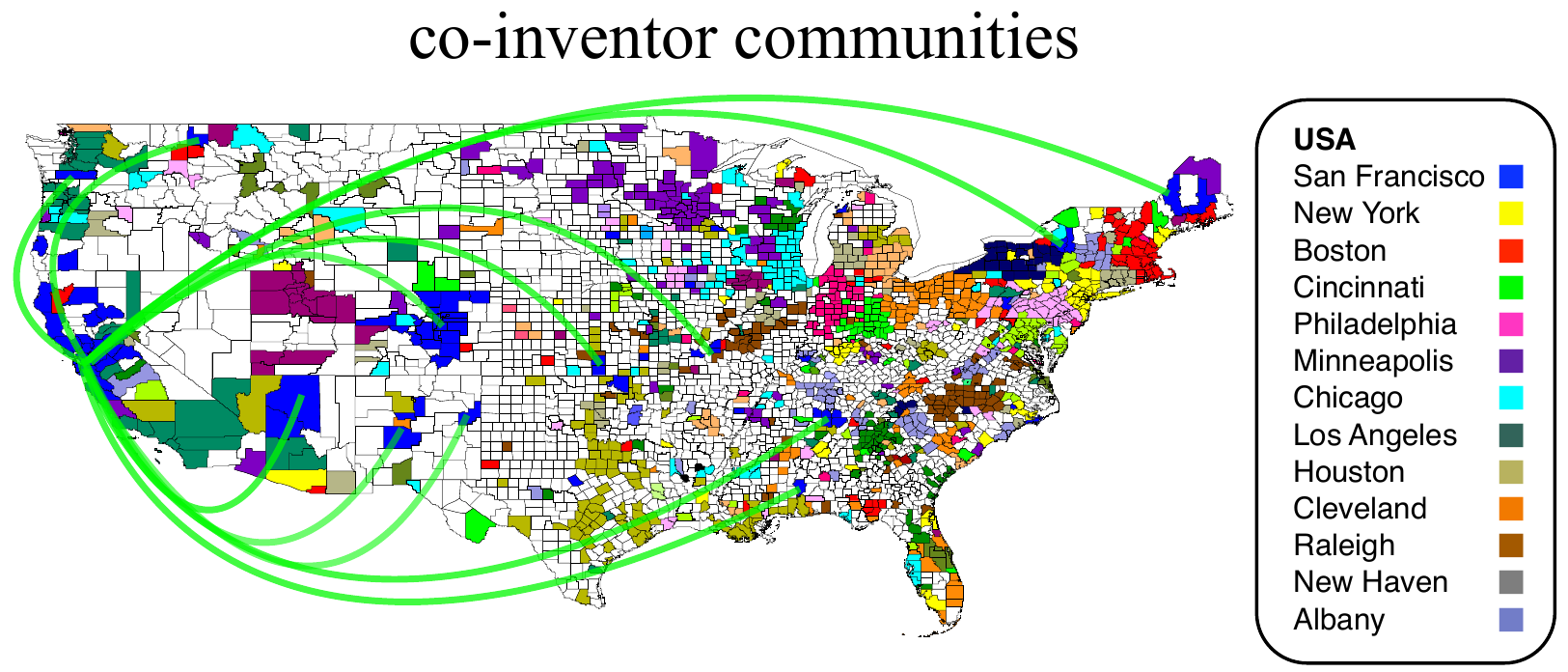}}
\caption{Community structure of the 2009 USA co-inventor network. We show only the top 13 communities and left regions belonging to all other communities white. The most central region of each community is listed in the legend and is determined by the procedure described in Section 2. Communities were determined using the Newman Girvan algorithm (21) and the Louvain algorithm (23). The green arcs are used to highlight some of the long distance members of the community for which San Francisco is the core region. Source: our computations based on data and code available here: {\tt http://cse.lab.imtlucca.it/SOM/SOM.zip}}
\label{FigS2}
\end{figure*}

After determining the community structure we calculated the centrality
 of each node within a community using a novel perturbative approach.
 Since we obtain the modularity score of a network
 ($Q$) by an optimization procedure, every perturbation of
 the partition structure leads to a negative variation in the
 modularity ($dQ$). For every node we calculate a $dQ$ by moving the node into every other community in the
 network. Within a specific community, the node with the most negative $dQ$ is
 defined as the most central node (core region). The legends of Fig. \ref{Fig1} in the manuscript and Fig. \ref{FigS2} below
 identify the most central nodes (using the city name associated with the
 NUTS3 region) for the top 13 communities in the 2009 co-inventor
 networks, for Europe and the USA respectively.

The community structure of the top 13 communities in the USA co-inventor network in 2009 is shown in Fig. \ref{FigS2}. Green arcs have been added to highlight (some of) the long range connections of the community for which San Francisco is the most central region. Communities in the USA have a higher fraction of cross border links than in the EU, indicating drastically different levels of integration in their respective R\&D collaboration networks. This result is confirmed numerically by comparing the share of links with at least one region outside of the nation in which the core region is located for Europe versus the share of links with at least on member outside the state in which the core region is located for US communities. This share is always significantly higher in the US co-inventor than in the EU. 
Table \ref{T2} shows that the fraction of cross-border ties is on average   larger for the US ($0.706$) as compared to  the EU ($0.138$), with $t$-statistic 7.16 ($p < 0.01$). The only European cluster which has a cross-border connectivity comparable with the US ones is the Nordic cluster centered on Copenhagen, which has many members outside Denmark in Sweden and Finland.

\subsection{Statistical analysis}

The rate at which EU (NUTS3) regions are linking to regions in other EU countries is increasing due to two types of factors: those that are global
and those that are EU specific. Thus, to capture the effect of EU specific institutional factors we must account for the net effect of the
global factors. In technical terms, we use the non-EU OECD members as a control group and its behaviour serves as the counterfactual
behavior of EU regions.\footnote{Given a general model with two state indicators A and B and two periods such as
\begin{equation*}
y =  \beta_{0}+\beta_{1}dA+\beta_{2}dB + \beta_{3}dA*dB + \delta_{0}d2 + \delta_{1}d2*dA + \delta_{2}d2*dB + \delta_{3}d2*dA*dB +u,
\end{equation*}
it can be easily shown that in a linear setting with no further explanatory variables the OLS estimate of the coefficient of the triple interaction term is just
\begin{equation*}
\hat{\delta}_{3}=(\bar{y}_{A,B,2}-\bar{y}_{A,B,1})-(\bar{y}_{N\!A,B,2}-\bar{y}_{N\!A,B,1})-(\bar{y}_{A,N\!B,2}-\bar{y}_{A,N\!B,1}),
\end{equation*}
where $N\!A$ and $N\!B$ indicate respectively the states not in A or not in B \cite{Wooldridge2010}. Underlying this analysis is the way we model the process that generates the link counts ($y_{i}$).
}

In our statistical analysis the number of links ($y_{i} \equiv A_{k,l}$) between NUTS3 regions $k$ and $l$
is regressed on a set of independent variables. We model this dependent variable with a count density.
A number of models can be found in the literature to handle count densities, including the Poisson model, Negative Binomial model variants,
and  Zero-inflated models \cite{PondsVanOortFrenken2007,ScherngellBarber2009,DeathofDist2009,HoekmanFrenkenTijssen2010,ScherngellBarber2011,HoekmanScherngellFrenkenTijssen2012,IntegrationERA}.
Since $\sim90\%$ of our link counts are zero, we opted for a zero-inflated negative binomial (ZINB),
as consistent with \cite{Wooldridge2010,ScherngellBarber2009}.\footnote{In the case of the inventor mobility network (and only that case)
the number of non-zero link counts was too low to be modeled using ZINB. Rather than tinkering with the threshold,
we modeled only the pairs of regions with $y_i>0$. A Zero Truncated Poisson model was employed in this special case.}
Zero-inflated models allow zeros to be generated by two distinct processes and are generally used when data exhibits ``excess zeros''
\cite{CameronTrivedi1998}. The ZINB model supplements a count density, $\tilde P$ with a binary zero generating process $\psi$. This allows a zero
count to be produced in two ways, either as an outcome of the zero generating process with probability $\psi$, or as an outcome of the count process
$\tilde P$ provided the zero generating process did not produce a zero ($\psi_{i}=1$).

The density distribution for the count pair  $y_{i}$ is then given by
\begin{equation}\label{eq:ZINB}
P(y_{i})=(1-\psi_{i})* \tilde P(y_{i}),
\end{equation}
where the zero generating process $\psi_{i}$ is parameterized as a logistic function of the regressors in $Z_{i}$, with parameter vector $\beta^{0}$:
\begin{equation}\label{eq:psi}
\psi_{i}=\frac{\exp(Z_{i}\beta^{0})}{1+\exp(Z_{i} \beta^{0})}.
\end{equation}
The count process $\tilde P(y_{i})$ is modeled as Negative Binomial of the second kind (NB2):
\begin{equation}\label{eq:Ppsi}
\tilde P(y_{i})=\frac{\Gamma(y_{i}+\alpha^{-1})}{\Gamma(y_{i}+1)+\Gamma(\alpha^{-1})} \left(\frac{\alpha^{-1}}{\alpha^{-1}+\mu_{i}}\right)^{\alpha^{-1}} \left(\frac{\mu_{i}}{\alpha^{-1}+\mu_{i}}\right)^{y_{i}},
\end{equation}
where the conditional mean $\mu_{i}$ is parameterized as an exponential function of the linear index $X \beta^{1}$, and $\alpha (\geq 0)$ is the
overdispersion parameter. Thus, drawing together equations \ref{eq:ZINB}, \ref{eq:psi}, and \ref{eq:Ppsi} our model for the expected count is
\begin{equation}
E(y_{i}|X_{i},Z_{i})=\left(1-\frac{\exp(Z_{i}\beta^{0})}{1+\exp(Z_{i} \beta^{0})}\right)*\exp(X_{i} \beta^{1}).
\end{equation}
In our estimation procedure we assume $X_{i}=Z_{i}$ because there is no reason to expect some variables would be relevant only in one of the two processes. However, individual regressors can impact the $y_{i}$ estimator
differently through the two distinct processes and their separate parameter vectors, $\beta^{0}$ and $\beta^{1}$.

The linear indices $X \beta^{0}$ for the zero-generating process and $X \beta^{1}$ for the Negative Binomial process are modeled in parallel as
\begin{equation}\label{eq_xb}
\begin{split}
X\beta^{j} & =  \beta_{0}^{j}+\beta_{1}^{j}border+\beta_{2}^{j}eu + \beta_{3}^{j}Distance + \beta_{4}^{j}Size_{k} + \beta_{5}^{j}Size_{l} + \gamma^{j} border*eu +\\
& + \sum_{t=2}^{T}  \theta_{t}^{j} year_{t} +\sum_{t=2}^{T} \delta_{t}^{j} border*year_{t} +\sum_{t=2}^{T} \zeta_{t}^{j} eu*year_{t} +\sum_{t=2}^{T} \eta_{t}^{j} border*eu*year_{t},
\end{split}
\end{equation}
where $j=0,1$. 
$Size_{k}$ and $Size_{l}$ denote the size of each of the two regions. We proxy the size of a region by the total number of links attached to the region. $Distance$ is the geographical
distance between centroids of the regions and $year_{t}$ is the year dummy variable. The time interval for estimation is generally 1986-2010 for patents and 1991-2009 for publications.
$border$ flags pairs of NUTS3 regions
within the same country. $eu$ flags pairs of NUTS3 regions that are within the EU ($eu=1$) and pairs of NUTS3 regions
for which neither are in the EU ($eu=0$). \footnote{Note that this dummy variable ($eu$) does not account for pairs of regions for which one is in the EU and one is not. Such links are not included as they are simply not relevant to the comparison we are focusing on.} Cross-sections are pooled over years and estimation is carried out on the whole sample clustering standard errors at pairs of NUTS3 regions. Following the Difference-in-Differences (DiD) econometric strategy, the full set of double/triple interaction dummy variables among the three dimensions ($eu=\{0,1\}$, $border=\{0,1\}$, $year_{t}=\{0,1\}$ for
$t=2,\ldots,T$) is relevant to the identification of treatment effect.

In the literature on program evaluation, DiD estimation is one of the most popular strategies for identifying the impact of a policy or
treatment  \cite{AngristKrueger1999,HeckmanLalondeSmith1999,AtheyImbens2006,BlundellCostaDias2009}. Treatment effect on an outcome variable is, in general,
defined as the difference between the outcome actually observed under the treatment and the counterfactual, that is the outcome
that would have been observed without treatment \cite{AtheyImbens2006}.
Under this treatment-effect framework,
our analysis seeks to quantify the effect of EU institutional changes upon integration {\it within} the EU, by measuring the relative rate of cross-border
links within a given network. Moreover, to isolate the signal arising only from EU factors we must control for the global rate of cross-border integration.
Specifically, we extend the standard DiD strategy of one state indicator (treatment vs control group) to the case of two state indicators, providing a control
group of links between non-European countries. For the purpose of embedding the institutional comparison in a temporal perspective, our analysis also includes a
time variable, in line with the standard treatment-effect formalism. Due to the addition of a second state indicator
our approach is a Difference-in-Differences-in-Differences estimator (DiDiD) \cite{Wooldridge2010}.

While the standard version of DiD estimation is designed for the linear case, it can be extended to cases in which the outcome variable is non-continuous and nonlinear estimation is preferred, as in our case \cite{Puhani2012}.  When employing the ZINB model it is not possible to make a general statement regarding the sign of the treatment effect merely by checking the sign of the interaction term(s) coefficients.
However, we can identify the treatment effect by calculating the incremental effect of the interaction term through comparison of a given year to a baseline year. In our framework, treatment effects are incremental effects of the triple interaction terms $border*eu*year_{t}$, evaluated at means of the regressors.\footnote{See \cite{Winkelmann2008} for the computation of marginal effects for the ZINB model.}

Denoting the actual and counterfactual
outcomes of our count dependent variable as $Y^{1}$ and $Y^{0}$ respectively and taking into account our DiDiD extension, the yearly treatment effect ($\tau_{t}$) can be defined as
\begin{equation}\label{eq_tau}
\begin{split}
\tau_{t}(year_{t}=1, eu=1, border=1, M) = & E[Y^{1}|year_{t}=1, eu=1, border=1, M] \\
& -E[Y^{0}|year_{t}=1, eu=1, border=1, M],
\end{split}
\end{equation}
where $M$ is the matrix of controls ($Size_{k}, Size_{l}, Distance$). Given the linear indices modeled in Eq. \ref{eq_xb} the expectation values of $Y^{1}$ and $Y^{0}$, for the group under treatment are
\begin{equation} \label{eq_Y1}
\begin{split}
E[Y^{1}|year_{t}=1, & eu=1,border=1,M]= \\
& \frac{\exp(\phi^{1}+\omega^{1}M+\eta_{t}^{1})}
{1+\exp(\phi^{0}+\omega^{0}M+\eta_{t}^{0})},
\end{split}
\end{equation}
\begin{equation} \label{eq_Y0}
\begin{split}
E[Y^{1}|year_{t}=1, & eu=1,border=1,M]= \\
& \frac{\exp(\phi^{1}+\omega^{1}M)}
{1+\exp(\phi^{0}+\omega^{0}M)},
\end{split}
\end{equation}
where $\phi^j \equiv \beta_{0}^{j}+\beta_{1}^{j}+\beta_{2}^{j}+\gamma^{j}+\theta_{t}^{j}+\delta_{t}^{j}+\zeta_{t}^{j}$, and
where $\omega^{1}$ and $\omega^{0}$ are the coefficient vectors for $M$ (the controls). 
We can then write Eq. \ref{eq_tau} as
\begin{equation}\label{eq_tau2}
\tau_{t}(year_{t}=1, eu=1, border=1, M) = \frac{\exp(\phi^{1}+\omega^{1}M+\eta_{t}^{1})}{1+\exp(\phi^{0}+\omega^{0}M+\eta_{t}^{0})}-
\frac{\exp(\phi^{1}+\omega^{1}M)}{1+\exp(\phi^{0}+\omega^{0}M)}.
\end{equation}

 We calculate the yearly treatment effect $\tau_t$  in Eq. \ref{eq_tau2} using parameter estimates of $\beta^0$ and $\beta^1$ and the sample mean of $M$. Estimates for the parameters in vectors $\beta^0$ and $\beta^1$ are obtained through maximum likelihood. Although differences in the linear indices ($\eta_{t}^{1}$ and $\eta_{t}^{0}$) are constant across individuals for a given t, it is clear from  Eq. \ref{eq_tau2} that differences in the dependent variable depend on the values chosen for $M$.

Relative to the baseline year $t^{*}$ (we use the year 2003 in our analysis as indicated by red dots in Fig. \ref{Fig2} in the manuscript), the yearly treatment effect reflects the impact of changes in institutional factors specific to the EU which have taken place in a given year $t$. Estimates of $\tau$, which are
just marginal effects of the triple interaction term $border*eu*year_{t}$. Estimates are obtained averaging over all the variables in the model and thus refer to an ``average'' pair of regions. 


Due to the large number of zero entries, in the regression analysis we ignore regions with fewer than 50 total patents. For inventor mobility, the analysis focuses only on NUTS3 region pairs with at least one link (nonzero counts). Since in co-authorship network the fraction of zeros is lower, we do not use any cutoff.\footnote{Additional results at the more aggregate level of NUTS2 Regions and with different cutoffs have been produced and confirm our main findings. In other words, our results do not critically depend on the definition and level of aggregation of administrative regions we consider. They are made available upon request.}



\clearpage
\newpage

\addtolength{\tabcolsep}{-3.5pt}
\begin{landscape}
\begin{table}[htbp]
\scriptsize
  \centering
  \caption{ \label{linkcounts} Network data summary. Total numbers of intra-border and cross-border links calculated for each area by networks. Source: our computations based on data and code available here: {\tt http://cse.lab.imtlucca.it/SOM/SOM.zip}}
    \begin{tabular}{rrrrrrrrrrrrrrrrrrrrr}
    \addlinespace
    \toprule
         & \multicolumn{4}{c}{REGPAT} & \multicolumn{4}{c}{REGPAT} &
\multicolumn{4}{c}{REGPAT} & \multicolumn{4}{c}{REGPAT} &
\multicolumn{4}{c}{ISI Publications} \\
    \midrule
    \multicolumn{1}{c}{} & \multicolumn{4}{c}{Coinventor} &
\multicolumn{4}{c}{Coapplicant} & \multicolumn{4}{c}{Citations} &
\multicolumn{4}{c}{Mobility} & \multicolumn{4}{c}{Coauthor} \\
    \multicolumn{1}{c}{} & \multicolumn{2}{c}{EU links} &
\multicolumn{2}{c}{non-EU links} & \multicolumn{2}{c}{EU links} &
\multicolumn{2}{c}{non-EU links} & \multicolumn{2}{c}{EU links} &
\multicolumn{2}{c}{non-EU links} & \multicolumn{2}{c}{EU links} &
\multicolumn{2}{c}{non-EU links} & \multicolumn{2}{c}{EU links} &
\multicolumn{2}{c}{non-EU links} \\
    \multicolumn{1}{c}{} & \multicolumn{1}{c}{\multirow{2}[0]{*}{Intra }} & \multicolumn{1}{c}{\multirow{2}[0]{*}{Cross}} & 
\multicolumn{1}{c}{\multirow{2}[0]{*}{Intra }} &
\multicolumn{1}{c}{\multirow{2}[0]{*}{Cross }} &
\multicolumn{1}{c}{\multirow{2}[0]{*}{Intra }} &
\multicolumn{1}{c}{\multirow{2}[0]{*}{Cross }} &
\multicolumn{1}{c}{\multirow{2}[0]{*}{Intra }} &
\multicolumn{1}{c}{\multirow{2}[0]{*}{Cross }} &
\multicolumn{1}{c}{\multirow{2}[0]{*}{Intra }} &
\multicolumn{1}{c}{\multirow{2}[0]{*}{Cross }} &
\multicolumn{1}{c}{\multirow{2}[0]{*}{Intra }} &
\multicolumn{1}{c}{\multirow{2}[0]{*}{Cross }} &
\multicolumn{1}{c}{\multirow{2}[0]{*}{Intra }} &
\multicolumn{1}{c}{\multirow{2}[0]{*}{Cross }} &
\multicolumn{1}{c}{\multirow{2}[0]{*}{Intra }} &
\multicolumn{1}{c}{\multirow{2}[0]{*}{Cross }} &
\multicolumn{1}{c}{\multirow{2}[0]{*}{Intra }} &
\multicolumn{1}{c}{\multirow{2}[0]{*}{Cross }} &
\multicolumn{1}{c}{\multirow{2}[0]{*}{Intra }} &
\multicolumn{1}{c}{\multirow{2}[0]{*}{Cross }} \\
    \multicolumn{1}{c}{} & \multicolumn{1}{c}{} & \multicolumn{1}{c}{}
& \multicolumn{1}{c}{} & \multicolumn{1}{c}{} & \multicolumn{1}{c}{} &
\multicolumn{1}{c}{} & \multicolumn{1}{c}{} & \multicolumn{1}{c}{} &
\multicolumn{1}{c}{} & \multicolumn{1}{c}{} & \multicolumn{1}{c}{} &
\multicolumn{1}{c}{} & \multicolumn{1}{c}{} & \multicolumn{1}{c}{} &
\multicolumn{1}{c}{} & \multicolumn{1}{c}{} & \multicolumn{1}{c}{} &
\multicolumn{1}{c}{} & \multicolumn{1}{c}{} & \multicolumn{1}{c}{} \\
    \multicolumn{1}{c}{1986} & 34,827 & 859  & 55,153 & 511  & 1,415 &
915  & 1,216 & 42   & 31,567 & 11,643 & 52,205 & 16,437 & 1,605 & 140
& 2,099 & 75   &      &      &      &  \\
    \multicolumn{1}{c}{1987} & 38,526 & 1,077 & 62,617 & 595  & 1,569
& 884  & 1,390 & 38   & 35,150 & 15,371 & 60,750 & 21,111 & 1,807 &
171  & 2,503 & 95   &      &      &      &  \\
    \multicolumn{1}{c}{1988} & 43,512 & 1,325 & 75,417 & 790  & 1,701
& 1,301 & 1,746 & 84   & 42,843 & 19,359 & 81,566 & 27,388 & 2,083 &
170  & 3,255 & 143  &      &      &      &  \\
    \multicolumn{1}{c}{1989} & 50,462 & 1,596 & 91,458 & 1,005 & 1,869
& 1,456 & 1,870 & 94   & 51,360 & 23,520 & 101,107 & 33,281 & 2,666 &
222  & 3,824 & 147  &      &      &      &  \\
    \multicolumn{1}{c}{1990} & 54,636 & 2,095 & 105,193 & 1,216 &
1,966 & 1,305 & 2,278 & 93   & 55,072 & 26,822 & 121,120 & 40,673 &
2,909 & 253  & 5,586 & 198  &      &      &      &  \\
    \multicolumn{1}{c}{1991} & 47,755 & 1,521 & 104,199 & 1,246 &
1,618 & 1,141 & 2,139 & 91   & 55,263 & 26,743 & 134,172 & 45,161 &
2,824 & 272  & 5,860 & 250  & 2,112 & 582  & 10,230 & 774 \\
    \multicolumn{1}{c}{1992} & 54,524 & 2,164 & 112,028 & 1,687 &
1,594 & 1,322 & 2,252 & 152  & 55,367 & 27,424 & 150,384 & 49,265 &
3,204 & 309  & 6,061 & 321  & 3,519 & 1,674 & 11,376 & 1,115 \\
    \multicolumn{1}{c}{1993} & 53,864 & 1,933 & 109,058 & 1,818 &
1,373 & 907  & 2,070 & 187  & 57,098 & 28,976 & 149,610 & 47,845 &
3,096 & 314  & 5,930 & 311  & 3,792 & 2,296 & 13,376 & 1,495 \\
    \multicolumn{1}{c}{1994} & 57,712 & 2,328 & 117,467 & 2,070 &
1,644 & 981  & 2,260 & 164  & 54,747 & 27,143 & 146,215 & 53,723 &
3,352 & 341  & 6,700 & 388  & 4,850 & 1,856 & 16,324 & 2,205 \\
    \multicolumn{1}{c}{1995} & 68,499 & 3,014 & 135,726 & 3,218 &
1,994 & 1,050 & 2,719 & 246  & 51,941 & 28,434 & 151,797 & 51,539 &
3,903 & 451  & 7,791 & 511  & 3,820 & 1,808 & 17,727 & 2,054 \\
    \multicolumn{1}{c}{1996} & 71,140 & 3,465 & 143,689 & 3,147 &
1,850 & 934  & 2,772 & 225  & 59,429 & 27,847 & 171,307 & 56,276 &
4,015 & 480  & 8,575 & 614  & 5,074 & 2,674 & 20,550 & 2,668 \\
    \multicolumn{1}{c}{1997} & 84,402 & 4,464 & 165,497 & 4,271 &
2,247 & 881  & 2,965 & 251  & 51,943 & 28,495 & 163,503 & 53,387 &
4,805 & 604  & 9,409 & 677  & 7,571 & 2,962 & 25,651 & 3,336 \\
    \multicolumn{1}{c}{1998} & 95,005 & 4,933 & 189,313 & 4,135 &
2,422 & 745  & 3,516 & 366  & 42,487 & 27,389 & 176,398 & 59,144 &
5,217 & 674  & 10,342 & 842  & 8,929 & 3,561 & 42,094 & 4,913 \\
    \multicolumn{1}{c}{1999} & 105,492 & 6,611 & 206,325 & 5,226 &
2,892 & 748  & 3,658 & 362  & 42,202 & 27,138 & 166,953 & 57,620 &
5,739 & 802  & 11,418 & 898  & 13,434 & 5,187 & 41,402 & 4,595 \\
    \multicolumn{1}{c}{2000} & 116,113 & 7,869 & 236,805 & 6,585 &
2,896 & 775  & 4,587 & 383  & 42,440 & 28,039 & 181,899 & 60,933 &
6,485 & 859  & 13,086 & 1,128 & 17,210 & 6,627 & 48,664 & 6,085 \\
    \multicolumn{1}{c}{2001} & 131,038 & 8,827 & 289,700 & 6,253 &
2,795 & 877  & 5,600 & 476  & 41,729 & 25,933 & 184,064 & 65,369 &
6,935 & 1,035 & 15,080 & 1,283 & 18,558 & 8,514 & 45,597 & 7,226 \\
    \multicolumn{1}{c}{2002} & 137,527 & 8,522 & 305,854 & 7,862 &
3,071 & 1,027 & 4,988 & 377  & 39,678 & 25,410 & 189,116 & 67,793 &
7,122 & 1,114 & 14,675 & 1,495 & 18,146 & 7,740 & 46,694 & 6,944 \\
    \multicolumn{1}{c}{2003} & 140,980 & 9,900 & 295,097 & 8,549 &
2,768 & 1,131 & 5,242 & 478  & 41,395 & 26,393 & 167,170 & 69,722 &
7,078 & 1,223 & 15,783 & 1,621 & 19,363 & 7,384 & 54,604 & 8,604 \\
    \multicolumn{1}{c}{2004} & 144,498 & 10,751 & 309,835 & 9,632 &
3,090 & 1,170 & 6,192 & 502  & 40,965 & 26,125 & 178,485 & 75,453 &
7,385 & 1,285 & 16,864 & 1,875 & 21,376 & 10,215 & 57,533 & 8,673 \\
    \multicolumn{1}{c}{2005} & 160,262 & 11,195 & 331,276 & 10,365 &
6,653 & 1,044 & 6,118 & 529  & 40,396 & 25,370 & 165,813 & 73,323 &
7,889 & 1,384 & 17,210 & 2,005 & 22,622 & 10,623 & 55,137 & 10,325 \\
    \multicolumn{1}{c}{2006} & 160,272 & 11,447 & 346,867 & 10,916 &
3,108 & 1,130 & 6,307 & 647  & 51,257 & 30,409 & 172,981 & 78,750 &
7,948 & 1,574 & 18,012 & 2,264 & 21,645 & 10,753 & 56,450 & 10,179 \\
    \multicolumn{1}{c}{2007} & 167,742 & 12,172 & 337,697 & 11,421 &
3,253 & 1,242 & 5,819 & 657  & 52,246 & 31,290 & 166,443 & 75,349 &
8,440 & 1,679 & 18,211 & 2,207 & 24,671 & 13,815 & 62,971 & 14,263 \\
    \multicolumn{1}{c}{2008} & 164,220 & 11,789 & 329,244 & 11,257 &
3,903 & 1,251 & 6,135 & 682  & 54,673 & 35,410 & 149,728 & 74,718 &
8,479 & 1,710 & 18,430 & 2,410 & 23,850 & 17,171 & 68,754 & 22,868 \\
    \multicolumn{1}{c}{2009} & 156,698 & 11,986 & 315,711 & 10,540 &
3,922 & 1,216 & 5,796 & 672  & 75,729 & 43,273 & 124,083 & 63,374 &
8,259 & 1,627 & 17,974 & 2,485 & 27,602 & 25,660 & 64,837 & 17,420 \\
    \multicolumn{1}{c}{2010} & 85,916 & 7,695 & 205,397 & 7,752 &
1,776 & 464  & 2,665 & 481  & 78,357 & 52,762 & 107,040 & 60,059 &
4,886 & 930  & 10,945 & 1,329 &      &      &      &  \\
    \multicolumn{1}{c}{Total} & 2,425,622 & 149,538 & 4,976,623 &
132,067 & 63,389 & 25,897 & 92,300 & 8,279 & 1,245,334 & 696,718 &
3,613,909 & 1,377,693 & 128,131 & 19,623 & 265,623 & 25,572 & 268,144
& 141,102 & 759,971 & 135,742 \\
    \bottomrule
    \end{tabular}%
  \label{tab:addlabel}%
\end{table}%
\end{landscape}

\begin{table}[htbp]
  \centering
  \caption{ \label{T2} Measurement of cross-border share of communities found in the coinventor networks for Europe and the USA in 2009. Data are the same as in Figs. \ref{Fig1} and \ref{FigS2}. The $t$-statistic is computed on the difference between averages. For the average percent multistate value $0.706, t = 7.16 (P < 0.01)$.}
    \begin{tabular}{rrrrrrrrr}
    \addlinespace
    \toprule
         & Intra- & Multi- & Percent &      &      & Intra- & Multi- &
Percent \\
         & country & country & multi- &      &      & state & state & multi- \\
         & links & links & country &      &      & links & links & state \\
             \hline
    \hline
    Mannheim & 1,612 & 12   & 0.007 &      & San Francisco & 208  &
396  & 0.656 \\
    Dusseldorf & 1,484 & 16   & 0.011 &      & New York & 178  & 584  & 0.766 \\
    Paris & 2,060 & 26   & 0.012 &      & Boston & 66   & 384  & 0.853 \\
    Hamburg & 2,218 & 6    & 0.003 &      & Cincinnati & 98   & 168  & 0.632 \\
    Stuttgart & 672  & 2    & 0.003 &      & Philadelphia & 148  & 258
 & 0.635 \\
    Eindhoven & 742  & 814  & 0.523 &      & Minneapolis & 232  & 328
& 0.586 \\
    Munich & 856  & 0    & 0.000 &      & Chicago & 98   & 428  & 0.814 \\
    Cambridge & 2,060 & 124  & 0.057 &      & Los Angeles & 54   & 324
 & 0.857 \\
    Copenhagen & 84   & 406  & 0.829 &      & Houston & 278  & 312  & 0.529 \\
    Nuremberg & 646  & 4    & 0.006 &      & Cleveland & 206  & 332  & 0.617 \\
    Milan & 1,184 & 2    & 0.002 &      & Raleigh & 178  & 254  & 0.588 \\
    Vienna & 392  & 14   & 0.034 &      & New Haven & 52   & 176  & 0.772 \\
    Madrid & 152  & 66   & 0.303 &      & Albany & 44   & 318  & 0.878 \\
    \hline
    \hline
    Average & 1,089 & 115 & 0.138 &      &      & 142 &
328 & 0.706 \\
    \bottomrule
    \end{tabular}%
  \label{tab:addlabel}%
\end{table}%

\clearpage
\newpage

\end{document}